\pdfoutput=1
\documentclass{article}

\usepackage{arxiv}
\usepackage[utf8]{inputenc} 
\usepackage[T1]{fontenc}    
\usepackage[disable]{todonotes} 
\usepackage{pythonhighlight}
\usepackage[frozencache,cachedir=.]{minted}
\usepackage{fancyvrb,fvextra}
\usepackage{geometry}
\usepackage{tcolorbox}
\usepackage{multirow}
\usepackage{array}
\usepackage{natbib}
\usepackage{booktabs}
\usepackage{doi}
\usepackage{graphicx}
\usepackage{amsfonts} 
\usepackage{hyperref}
\usepackage{url}
\usepackage{threeparttable}
\usepackage{footmisc} 

\DeclareUnicodeCharacter{180E}{~}
\DeclareUnicodeCharacter{200B}{~}

\definecolor{darkcandyapplered}{rgb}{0.64, 0.0, 0.0}
\definecolor{darkbyzantium}{rgb}{0.36, 0.22, 0.33}
\definecolor{falured}{rgb}{0.5, 0.09, 0.09}
\definecolor{frenchrose}{rgb}{0.96, 0.29, 0.54}
\definecolor{indianred}{rgb}{0.8, 0.36, 0.36}
\definecolor{jasper}{rgb}{0.84, 0.23, 0.24}
\definecolor{jazzberryjam}{rgb}{0.65, 0.04, 0.37}
\definecolor{persianplum}{rgb}{0.44, 0.11, 0.11}
\definecolor{red-brown}{rgb}{0.65, 0.16, 0.16}
\definecolor{britishracinggreen}{rgb}{0.0, 0.26, 0.15}
\definecolor{cobalt}{rgb}{0.0, 0.28, 0.67}
\definecolor{coolblack}{rgb}{0.0, 0.18, 0.39}

\title{Automatic Generation of Programming Exercises and Code Explanations with Large Language Models}

\author{
    Sami Sarsa\\
	Aalto University\\
	\texttt{sami.sarsa@aalto.fi} \\
	\And
	{\hspace{1mm}Paul Denny} \\
	The University of Auckland\\
	\texttt{paul@cs.auckland.ac.nz} \\
	\And
	{\hspace{1mm}Arto Hellas} \\
	Aalto University\\
	\texttt{arto@hellas@aalto.fi} \\
	\And
	{\hspace{1mm}Juho Leinonen} \\
	Aalto University\\
	\texttt{juho.2.leinonen@aalto.fi} \\
}

\renewcommand{\undertitle}{Preprint, accepted in ICER'22}

\newcommand{\verbatimfont}[1]{\renewcommand{\verbatim@font}{\ttfamily#1}}

\hypersetup{
pdftitle={Automatic Generation of Programming Exercises and Code Explanations with OpenAI Codex},
pdfsubject={TODO,},
pdfauthor={Sami Sarsa, Paul Denny, Arto Hellas, Juho Leinonen},
pdfkeywords={Natural language generation, OpenAI Codex, GPT-3, CS1, Programming exercises, Code explanations, Robosourcing, Exercise generation, Resource generation, Automated feedback},
}

\begin{document}
\maketitle

\begin{abstract}


This article explores the natural language generation capabilities
of large language models with application to the production of two types of learning resources common in programming courses.
Using OpenAI Codex as the large language model, we create programming exercises (including sample solutions and test cases) and code explanations, assessing these qualitatively and quantitatively. Our results suggest that the majority of the automatically generated content is both novel and sensible, and in some cases ready to use as is. 
When creating exercises we find that it is remarkably easy to influence both the programming concepts and the contextual themes they contain, simply by supplying keywords as input to the model. 
Our analysis suggests that there is significant value in massive generative machine learning models as a tool for instructors, although there remains a need for some oversight to ensure the quality of the generated content before it is delivered to students. We further discuss the implications of OpenAI Codex and similar tools for introductory programming education and highlight future research streams that have the potential to improve the quality of the educational experience for both teachers and students alike.

\end{abstract}


\section{Introduction}


Creating an introductory programming course involves designing course materials, developing assignments, creating feedback opportunities, 
and planning out the flow of the course~\cite{falkner2019pedagogical}.  Providing opportunities for active learning -- that involves both doing and reflecting -- is especially important for learning programming \cite{sanders2017folk}. One particular approach which has gained popularity due to the wide availability of auto-grading tools that can generate immediate feedback is the use of many short programming exercises which students use to develop mastery through regular practice~\cite{allen2018weekly,edwards2020syntax,vihavainen2011extreme}.  Such approaches have become so popular that Finnie-Ansley et al.\@ suggest they may form a signature pedagogy within computing education \cite{finnie2022robots, shulman2005signature}. 

Developing suitable programming exercises involves multiple facets, including creating problem statements, sample code solutions and automated tests.   Creating a sufficient quantity of novel exercises to form a useful resource is a significant challenge for educators, and particularly difficult in an age where solutions to existing exercises are quickly published and shared online.  Indeed, writing good questions and good tests to verify them is a fundamental challenge \cite{wrenn2018who, lobb2016coderunner}.  It is thus not surprising that researchers have explored collaborative approaches to resource generation, such as crowdsourcing, but these are not free from problems as they suffer from limitations around student motivation and content quality \cite{sanders2013canterbury,denny2011codewrite,pirttinen2018crowdsourcing}.



Let us consider a possible problem statement for a programming exercise that could appear early in a typical CS2 course focused on object-oriented programming \cite{porter2018developing}.  In this example, the language is Python and the goal is to familiarize students with the basic concepts of attributes, methods and method calling, as well as to practice manipulating built-in dictionaries.

\begin{tcolorbox}
\begin{Verbatim}[fontsize=\small,breaklines]
Write a class called Fisherman that is initialized with a dictionary of fish, e.g. {"Salmon": 10, "Trout": 20, "Tuna": 5}. The class should have a method called catch, which takes in one parameter, a fish name, and returns the number of that fish caught. The class should also have a method called throw_away, which takes in a fish name, and decreases the number of that fish in the dictionary by 1.

As an example, the code:
fisherman = Fisherman({"Salmon": 10, "Trout": 20, "Tuna": 5})
fisherman.catch("Salmon")
fisherman.throw_away("Trout")
fisherman.catch("Tuna")
print(fisherman.fish)

should print out:
{"Salmon": 11, "Trout": 19, "Tuna": 6}
\end{Verbatim}
\end{tcolorbox}


A student that chooses to work on this programming exercise might produce a solution similar to the following.

\begin{tcolorbox}
\begin{minted}[fontsize=\small]{python}
class Fisherman():
  def __init__(self, fish):
    self.fish = fish

  def catch(self, fish):
    if fish in self.fish:
      self.fish[fish] += 1
    else:
      self.fish[fish] = 1

  def throw_away(self, fish):
    if fish in self.fish:
      self.fish[fish] -= 1
      if self.fish[fish] == 0:
        del self.fish[fish]
\end{minted}
\end{tcolorbox}


Once the student submits their solution, they would typically receive some kind of feedback from the auto-grader.  If there is an error in their code, the student may attempt a line-by-line walkthrough of their solution to help them locate a bug.  If they are stuck, they may reach out to a teaching assistant for help, in which case the assistant may ask the student to explain their code and jointly walk through it with them step by step.  This type of communication, similar to rubber-duck debugging, can be an effective way of exploring one's own understanding \cite{phillips2013making}.  If the student needs a great deal of help, the teaching assistant may even help generate an explanation for them.  A stepwise explanation of the above program might be as follows.

\begin{enumerate}
    \item We create a dictionary called fish where we'll store our fishes and their quantity.
    \item We create a class called Fisherman.
    \item We create a constructor for the class Fisherman where we initialize the fish dictionary.
    \item We create a method called catch which takes a fish as a parameter.
    \item If the fish is already in the fish dictionary, we increment its value by 1.
    \item If the fish is not in the fish dictionary, we add it and set its value to 1.
    \item We create a method called throw\_away which takes a fish as a parameter.
    \item If the fish is in the fish dictionary, we decrement its value by 1.
    \item If the fish's value reaches 0, we delete the fish from the dictionary.
\end{enumerate}



Generating these types of explanations and jointly helping students trace through their own code can take a considerable amount of time and effort for teachers.  This effort grows proportionally with the cohort size, as does the challenge of developing repositories of practice exercises that suit the varied learning needs of larger and more diverse cohorts.  This makes the idea of automatically generating these kinds of materials an exciting prospect.  As a matter of fact, \emph{the above programming exercise, its solution, and the code explanation were all generated automatically by OpenAI Codex}, which is a generative NLP model for creating code and code related texts.  In addition to the exercise, solution and code explanation, OpenAI Codex also generated a suite of test cases for the exercise that could be used to automatically verify attempted solutions.  We discuss the generation of test suites later in the paper. Given the known difficulties around maintaining the integrity of existing question banks \cite{albluwi2019plagiarism}, it is notable
that the problem description itself is, as far as we can tell, entirely novel.  At the time of writing, the problem description returns no relevant matches (using any combination of sentence fragments from the description) on search engines like Google or on websites like Chegg or StackOverflow that are frequently used by students to find solutions from problems statements.

Very recent work by Finnie-Ansley et al.\@ has explored the implications of OpenAI Codex on programming education, but from the perspective of assessing the accuracy of source code generated by the model to solve typical CS1 test and exam questions~\cite{finnie2022robots}.  This prior study focused primarily on the challenges that the technology poses to educators, including serious academic integrity issues, over-reliance by novices, and confusion caused by the generation of incorrect code or code with poor style.  In this work we explore the opportunities that are provided by this new technology.  Instead of focusing on the generation of source code, which is the most publicized functionality of OpenAI Codex, we primarily investigate the generation of natural language artefacts -- both programming exercises and explanations of code -- that may offer value to both instructors and students. 


\begin{enumerate}
    \item [\emph{RQ1}] To what extent are programming exercises created using OpenAI Codex sensible, novel, and readily applicable? 
    \item [\emph{RQ2}] How comprehensive and accurate are OpenAI Codex natural language explanations of code solutions to introductory programming exercises?
\end{enumerate}

\begin{figure}[t!]
\centering
\includegraphics[width=0.4\textwidth]{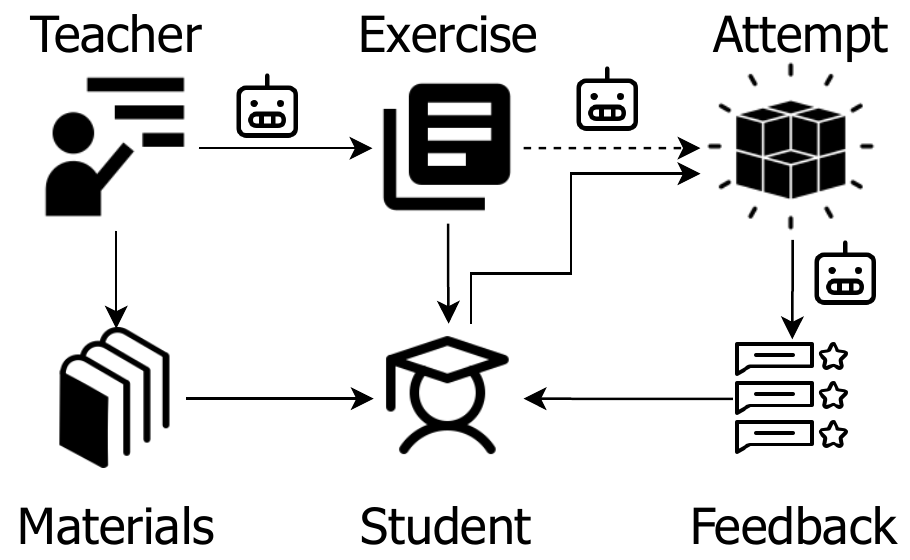}
\caption{Lifecycle of a programming exercise. Teacher creates programming exercises and learning materials. Students study the materials and the exercises, and create exercise attempts. Students receive feedback on their attempts. In this work, denoted by the solid arrows with a robot, we explore the use of OpenAI Codex for the creation of programming exercises and for providing feedback on students' programming exercise attempts. The dashed arrow with a robot represents prior work by Finnie-Ansley et al.\@~\protect\cite{finnie2022robots} who explored how well Codex can solve introductory programming exercises.\label{fig:course-design}}
\end{figure}

Our work provides insight into the utility of OpenAI Codex as one part of the toolbox of a teacher of an introductory programming course and discusses the further potential of such tools. In this work, we focus on the applicability of OpenAI Codex for the generation of programming exercises and for creating feedback from student attempts to programming exercises.  Figure ~\ref{fig:course-design}, which illustrates a simplistic lifecycle for a programming exercise, provides some context for our contributions.


\section{Background}

\subsection{Practice and Feedback in Introductory Programming Courses}

%



Introductory programming courses around the world interleave theory and practice, providing students opportunities for learning how to write programs guided by exercise statements, automated assessment systems, and course staff. Courses and course assignments are typically written so that they are increasingly complex, gradually introduce new concepts and seek to avoid overwhelming students, i.e.\@ seek to avoid cognitive overload~\cite{duran2021cognitive}. Such a design can be seen as scaffolding that supports students in their zone of proximal development~\cite{vygotsky1978mind}, that is, their area of skills and knowledge where they cannot yet succeed on their own, but where they can succeed with guidance. As a student learns, the student's zone of proximal development also changes.


The design of programming courses and course assignments often reflects the idea of deliberate practice~\cite{ericsson1993role}, which is a systematic and purposeful type of practice that focuses on improvement of performance in a specific task. Continuing deliberate practice is sustained with grit~\cite{duckworth2013true}, i.e.\@ passion and perseverance for long-term goals, even when pursuing those goals feels difficult. Motivation towards assignments is influenced by the design of assignments; while very easy assignments have a high expectancy for success, their utility value is low, and as per expectancy-value theory~\cite{rosenzweig2019expectancy}, students may have little motivation to work with them. Conversely, assignments that are too difficult also have little utility and can lead to low motivation~\cite{rosenzweig2019expectancy}, and likely contribute negatively towards feelings of self-efficacy~\cite{bandura1977self}. One practice in teaching introductory programming courses that has sought to avoid students prematurely encountering assignments that are too complex is the use of many small programming exercises which help develop mastery through regular practice~\cite{allen2018weekly,edwards2020syntax,vihavainen2011extreme}. As students have different backgrounds, different skills, and differently evolving zones of proximal development, each student would likely benefit from a tailored set of assignments, maybe even with contextual cues tuned to their own interests.  This latter point is supported by prior work in computing education suggesting that students' familiarity with the context of a problem can potentially be helpful~\cite{leinonen2021exploring}. However, such large exercise pools would be very tedious to develop~\cite{wrenn2018who, lobb2016coderunner}.



In addition to creating programming assignments, teachers often design feedback opportunities to course assignments. One common approach for providing feedback in introductory programming courses is the use of automated assessment systems~\cite{ihantola2010review,ala2005survey,paiva2022automated}, which at the minimum provide feedback on the correctness of programming assignments submitted for evaluation. As feedback plays a considerable role in learning~\cite{hattie2007power}, in addition to influencing approaches to learning by simply being offered~\cite{vollmeyer2005surprising}, it should be given with care; feedback can both improve self-efficacy and decrease self-efficacy~\cite{hattie2007power}. In general, formative feedback -- feedback given as a part of the learning process -- is preferred over summative feedback, i.e.\@ feedback given after the learning process~\cite{shute2008focus,keuning2018systematic}. In particular, formative feedback can be used to aid self-regulated learning and metacognition, helping students in becoming better learners~\cite{shute2008focus}. 

Classroom practices and the way that programming instruction is organized also matters~\cite{vihavainen2014systematic}. In particular, feedback opportunities can be included into classroom and lab sessions. For example, both peer instruction~\cite{crouch2001peer} and pair programming~\cite{williams2000strengthening} create opportunities for reflection. In peer instruction, the reflection is partially guided by the teacher responsible for the peer instruction questions, while in pair programming, students interact and reflect on the program that is being worked on. Students tend to enjoy pair programming~\cite{aarne2018study} and also learn to reason and explain code.




\subsection{Code Explanations and Their Assessment}
\label{background-code-explanation}

The ability to reason about code and explain its purpose is a key skill that novices develop as they gain expertise \cite{murphy2012ability}.  The relationship between a student's ability to explain code, and related skills such as code tracing and code writing, have been the focus of much prior research in computing education \cite{lister2006forest, lister2009further, venables2009closer, sheard2008going}.  The evidence from this body of work generally suggests that competence at explaining code develops after lower-level code tracing skills and before higher-level code writing skills.  This hierarchy forms the basis of a recently proposed theory of programming instruction by Xie et al.\@ in which learners first develop the ability to explain the purpose of reusable code templates before writing code to solve new problems \cite{xie2019theory}.  

Assessing both code tracing and code writing skills is generally straightforward as answers are objective and thus can be readily automated.  Code tracing questions typically ask students to determine the output or the value stored in a variable after a provided code block is executed and are often presented in multiple-choice format \cite{lister2004multinational}. Hassan and Zilles propose a novel `reverse-tracing' question format which is non-trivial even when students have access to a computer \cite{hassan2021exploring}, and Lehtinen et al.\@ explored automatically creating multiple choice questions from students' own code~\cite{lehtinen2021let}. A variety of approaches and tools for helping students develop code tracing skills have also been reported \cite{xie2018explicit, qi2020unlimited}.  Code writing questions require students to produce code from a problem description.  Many tools for automatically grading code writing questions have emerged as institutions shift away from paper-based exams \cite{nip2018using, stephenson2018experience}, and look to provide immediate feedback on programming assignments and tasks \cite{baniassad2021stop, ullah2018effect, leinonen2022comparison}. 

Code explanation skills are less straightforward to assess because explanations are usually given in natural language and can be provided at varying levels of abstraction.  A popular method for evaluating code explanation ability is the `explain in plain English' question format, where students are asked to explain the purpose of a provided code block.  This type of question was first studied as part of the BRACElet project \cite{whalley2006australasian}, where student responses were classified by researchers according to the first four levels of the SOLO taxonomy \cite{biggs1982SOLO}.  The highest of these levels, relational, characterized responses that described at a high-level of abstraction how the code would behave over all possible inputs.  A classic example from the BRACElet work is the response ``it checks to see if the array is sorted'' for describing code that compares adjacent elements in an array within a loop \cite{lister2020cognitive}.  The next highest level, multistructural, was used to classify responses that gave a line-by-line description of the code but failed to succinctly state its purpose.  Subsequent research has established a strong correlation between code writing skills and the ability to construct responses to explain in plain English questions at the relational level \cite{murphy2012ability}.  More recently, approaches for grading explain in plain English questions on exams have been explored, including a validated rubric to inform manual grading \cite{chen2020validated} and an automated tool which exhibited similar accuracy to that of trained teaching assistants \cite{fowler2021autograding}.  Both approaches, like the SOLO classification commonly used in research, differentiate between responses at an abstract level and those which are line-by-line descriptions of the code.

Clearly, the ability to explain the purpose of \emph{correct} code at an abstract level is an important skill for novices to develop.  However, when code is \emph{incorrect} the situation is more complex.  For code that contains bugs, attempts to describe its purpose at the relational level are premature and will likely not identify the errors. 
Indeed, Perkins et al.\@ argue that the ability to read what a piece of code actually does, rather than what we think it might do on a quick first inspection, is an important debugging skill \cite{perkins1986conditions}.  They describe line-by-line walkthroughs of code as `close tracking', which mirrors to some extent the `mental simulations' that Soloway argues should be taught explicitly to students \cite{soloway1986learning}. 
Therefore it is possible that multistructural explanations of code, at a line-by-line level, may provide some benefit for the purposes of debugging. 
Given that code walkthroughs can be mentally demanding, being presented with an explanation of one's own code may reduce the cognitive demands associated with debugging \cite{mccauley2008debugging}  
and evidence from other educational domains suggests that being presented with explanations produced by others can improve learning \cite{williams2016axis}.
Within computing education, techniques like pair programming \cite{hanks2011pair} and misconception-based peer feedback \cite{kennedy2020misconceptions} provide some opportunities for walking through code with others, but are not always feasible and may not be suitable for individual student assessments.  
Therefore, the automatic generation of code explanations, particularly for supporting student debugging, is an attractive idea and one which is made feasible with the introduction of tools like OpenAI Codex.

\subsection{Machine Learning Models for Code Generation}


Recently, there has been great progress on generative natural language models, such as OpenAI's GPT-3~\cite{brown2020language}, that are capable of generating text that can be hard to distinguish from text written by humans~\cite{brown2020language}. These are deep learning models and their performance relies on both a vast number of parameters for the models (175 billion in the case of GPT-3) as well as an extensive corpus of text for training (570GB of text for GPT-3). Codex~\cite{chen2021evaluating}, also by OpenAI, is a GPT-model similar to GPT-3 but has been fine-tuned using publicly available code from GitHub with the goal of translating natural language to source code and vice versa, and to generate or auto-complete source code given source code as input.

In addition to OpenAI's Codex, other generative machine learning models capable of generating natural language from source code and/or vice versa have been developed. One of the earliest such models is Microsoft's CodeBERT~\cite{feng2020codebert}. CodeBERT has been trained with natural language - programming language pairs and is capable of generating source code documentation automatically similar to Codex. Another recently presented model is DeepMind's AlphaCode~\cite{li2022competition}, which is capable of performing on par with a median competitor when presented with problem prompts at the programming competition level.

These recent models have multiple applications. One that has been proposed is to help programmers fix insecure code. Pearce et al.\@~\cite{pearce2021can} analyzed the performance of Codex and similar models for repairing source code containing security flaws and found that through providing a carefully constructed prompt for the model, they were able to patch security issues in programs in some cases. Another study by Pearce et al.\@~\cite{pearce2022pop} analyzed the possibility of utilizing Codex for reverse engineering. In their study, they provided Codex decompiled code and prompted Codex to explain the purpose of the code. Their results indicated that  there is some potential in utilizing models such as Codex for reverse engineering as slightly over half of the questions authors asked were answered correctly.   However, they suggest there is a need for ongoing work, and propose fine-tuning the model by providing it context-specific training data (in their case, decompiled source code).

Finally, relevant to the current paper, very recent work in the domain of mathematics has shown that large language models can successfully solve and generate new problems~\cite{drori2021machine}.  Interestingly, in order to solve the mathematics problems, the authors use Codex to generate code-based solutions from natural language prompts which are then executed in order to perform the calculations. 

\subsection{Potential of Codex in Computing Education}

The most common use case for Codex is generating new code from either a provided code fragment or from a natural language description of a problem.  GitHub Copilot, which is an editor plug-in that is powered by OpenAI Codex, promises to generate code suggestions ``for whole lines or entire functions right inside your editor''. Indeed, the tagline for Copilot is: ``your AI pair programmer''.  A developer using this plug-in would typically receive real-time code suggestions as they are typing, or would explicitly provide a natural language description (for example, as a code comment) and then receive more comprehensive suggestions, such as entire functions, almost immediately.  In many cases, multiple suggestions are provided which the developer can simply cycle through and accept or reject. 

This code-generation use case could be applied productively in several ways in computing education contexts.  For example, as model solutions have been proposed as a support mechanism in introductory programming~\cite{nygren2019experimenting,nygren2019non}, students could generate model solutions with Codex for historical assignment, test and exam problems, where solutions may not otherwise exist.  They could also generate alternative correct solutions for a problem they have solved, to reflect on their own solution and to compare different algorithms and language constructs.  As the accuracy of Codex improves over time, introductory computing pedagogy may shift away from low-level coding and towards problem decomposition and problem solving. 

However, Codex is not limited to code-generation tasks, and can generate natural language output from code-based or prose-based input.  In the current paper, we explore how this capability can be used to support two novel use cases that relate to the programming exercise lifecycle (see Figure~\ref{fig:course-design}).  The first of these relates to the generation of programming exercises by the instructor.  Given an existing exercise as input, we explore whether Codex can generate novel variations of the exercise that could then be deployed to students.  The second of these relates to the generation of feedback to students.  Given source code as input, we explore whether Codex can produce useful natural language feedback on that code, particularly in terms of helping students detect bugs prior to submission for grading.  In general, the use of a tool like Codex to generate practice problems for computing students in various formats, and to provide useful feedback to students on their progress on those problems, appears to offer great potential. 

In the context of programming education, Finnie-Ansley et al.\@~\cite{finnie2022robots} studied the potential of Codex for solving introductory programming assignments. They found that Codex was able to correctly answer most introductory programming problems and that when given typical exam questions, Codex performed better than the average student. The authors note that considering the performance of Codex, and especially that the progress in this area has been rapid, there are clear consequences for introductory programming courses. For example, when models such as Codex that are capable of performing well on programming assignments become more and more common, it becomes increasingly easy for students to use these models to write code for them, essentially engaging in a new type of plagiarism, which might require the utilization of process-based plagiarism detection~\cite{hellas2017plagiarism,longi2015identification,leinonen2016typing}. While Finnie-Ansley et al.\@ focused mostly on potential challenges Codex-like models will introduce to introductory programming classrooms, our focus in this article is exploring potential opportunities these models can provide in programming education.







\section{Methodology}

\subsection{Using Codex \label{subsec:using-codex}}

Similar to OpenAI's GPT-3 models, Codex can be used both programmatically through an API or through a web-UI. The user provides a priming, i.e.\@ a prompt, to Codex as input and Codex generates new content as output based on the given priming. For example, given a natural language description of desired behavior, Codex will often generate source code for a program that provides that functionality. 

For generating content, a custom ``stop sequence'' can be specified, which causes the generation of text to stop upon creating such a sequence. Other relevant options that we leveraged in this study include maximum token count that controls the length of the generated content and ``temperature'' that controls the ``creativity'' or ``randomness'' of the model. A lower temperature value will further reduce the chances of the model generating less probable tokens, reducing randomness in the creation process. With any temperature value, however, the model is not deterministic and there can be differences in the created content between runs, although this is more common with higher temperature values.







Since the priming given to Codex primes the model on what content should be generated, in addition to generating code, we can for instance prime Codex with an existing programming exercise and some context related words. This guides Codex to try and create content similar to the priming, which in this case would be a similar exercise but with a specified context. For reference, considering e.g.\@ the natural language model GPT-3 (which Codex is based on), using a priming about dogs will likely lead to output related to dogs. 

\subsection{Creating Programming Exercises and Code Explanations}


\subsubsection{Choosing inputs for Codex\label{subsubsec:choosing-inputs}}

For the purposes of the analyses in this article, we selected a small set of exercises that have been featured in computing education research and that are often used in the teaching contexts of the researchers, who use the many small exercises approach~\cite{allen2018weekly}. We focused on four programming exercises: 1) a variant of the speeding problem~\cite{venables2009closer} where students work with conditionals and returning values, 2) a variant of FizzBuzz~\cite{althoff2022self} where students work with conditionals and lists, and study the importance of ordering of conditional statements, 3) a variant of the Rainfall Problem~\cite{soloway1984empirical} that has been a recurring problem in computing education research~\cite{seppala2015we,fisler2014recurring}, and 4) a currency converter application used in our contexts where students work with objects, methods, and dictionaries. A sample solution for each of the four programming exercises is shown in Appendix~\ref{appendix:explanation-priming-codes}.

Since OpenAI Codex has primarily been evaluated with the Python programming language in prior work~\cite{chen2021evaluating} and reportedly works best in Python\footnote{As noted in the OpenAI Codex Beta documentation, last accessed 2022-03-25:\\ \url{https://beta.openai.com/docs/engines/codex-series-private-beta}}, all of our exercises that we use to prime OpenAI Codex are in Python. In our explorations, we used the \texttt{code-davinci-001} Codex model, which was the most capable (albeit slowest) version when these experiments were conducted.


\subsubsection{Creating programming exercises}

We explored a range of priming approaches for creating programming exercises. In the end, the priming that we found most reliable for creating new programming exercises contained a problem description, a sample solution, and automated tests. In addition, we explored adding programming-related concepts (e.g.\@ conditional, loop) and contextual concepts (e.g.\@ hiking, fishing) to the priming. In general, we observed that introducing concepts led to OpenAI Codex taking these into account when creating programming exercises, although the programming exercises created without the concepts were also meaningful.
To see how these primings are formatted, refer to Appendix~\ref{appendix:creation-priming-exercises}. In addition to the examples in Appendix~\ref{appendix:creation-priming-exercises}, when providing the samples as an input to Codex, the samples were suffixed with a stop sequence (\texttt{"""}). After the stop sequence, the priming included the text ``\texttt{Exercise 2}'', the concepts desired in the created exercise and the identifier for the problem statement (\texttt{-{}-Problem statement-{}-}). An example of a complete priming (i.e.\@ the input to Codex) and one example of the corresponding output generated by Codex can be found in Appendix \ref{appendix:creation-full-priming}. 


\begin{table*}[h!]
    \centering
    \caption{Keywords used for priming exercise generation. The programming-related concepts are placed in two sets to reduce the number of possible combinations. \label{tbl:priming-keywords}}
    \small
    \begin{tabular}{l|l|l}
    \toprule
    contextual concepts & programming-related concept set 1: ``function'' & programming-related concept set 2: ``class'' \\
    \midrule
    \multirow{5}{8em}{hiking,
        fishing,
        relationships,
        football,
        music,
        health,
        ice hockey,
        books,
        cooking
    }
    & function & class \\
    & parameters & list \\
    & dictionary & list comprehension \\
    & dict comprehension & conditional \\
    & arithmetics & \\
    \bottomrule 
    \end{tabular}
\end{table*}

We generated exercises using the two priming exercises in Appendix~\ref{appendix:creation-priming-exercises}, varying both the programming-related concepts and the contextual concepts (see Table~\ref{tbl:priming-keywords}). Using a total of nine contextual concepts (and an extra for leaving out the contextual concept) and two pro\-gram\-ming-related concept sets (and an extra for leaving out the programming-related concepts), we generated a total of $10 \times 3 \times 2 = 60$ different combinations of inputs (contextual concepts $\times$ programming-related concept sets $\times$ exercise primings). In addition, we explored two values for Codex's temperature parameter ($0$ and $0.75$) and created two exercises for each parameter combination. In total, this led to a sample of $60 \times 2 \times 2 = 240$ programming exercises.

\subsubsection{Creating code explanations}

Similar to creating programming exercises, we explored different types of priming approaches for creating code explanations. We identified three types of primings that led to different types of code descriptions: 1) a high-level description of the code, 2) a problem statement-like description of the code, and 3) a step-by-step explanation of the code. In this work, we focus on the last code explanation type, i.e.\@ the step-by-step explanation of code, as it aligns with the multistructural level of the SOLO taxonomy and is often produced by students when prompted to explain code~\cite{lister2006forest}.

In our experiments, using a priming that consisted of the source code, followed by a stop sequence and the text ``Step-by-step explanation of the above program:'', and a number one followed by a dot, tended to produce step-by-step explanations. As an example, the priming for a simple ``Hello world!'' program would look as follows:

\begin{tcolorbox}
\begin{Verbatim}[fontsize=\small,breaklines]
print("Hello world!")

"""Step-by-step explanation of the above program:
1.
\end{Verbatim}
\end{tcolorbox}

With the above priming, Codex would create a step-by-step explanation of the code \texttt{print("Hello world!")}. For the step-by-step code explanation analysis, we created five explanations for each of the four programming exercise sample solutions in Appendix~\ref{appendix:explanation-priming-codes}, leading to a total of $20$ code explanations. Since we were interested in precise explanations instead of creative ones, we used the temperature value $0$
to generate each of the explanations. 




\subsection{Evaluation}

\subsubsection{Programming exercises}

The evaluation of the programming exercises was conducted as mixed-methods research, where the exercises were evaluated both qualitatively and quantitatively.

In the qualitative analysis, we focused on a random sample of 120 programming exercises. Our focus was on the sensibleness, novelty and readiness for use of the created programming exercises, as outlined in RQ1. When assessing \emph{sensibleness}, we study whether the programming exercise represents a sensible problem for students -- does the problem statement describe a practical problem that could be given to students to solve? When assessing \emph{novelty}, we study whether the verbatim copy of the programming exercise or a similar programming exercise already exists and can be found online (we used both Google and GitHub for searching). Related to novelty, we also examine the \emph{topicality} of the exercises -- how are the different priming concepts accounted for in the created exercises? When assessing \emph{readiness for use}, we consider the amount of manual work a teacher would have to make to the exercises and the associated sample solution and tests.

The qualitative analysis was conducted by four researchers, who first practiced the assessment of sensibleness, novelty, and readiness for use jointly, discussing identified issues and corner cases. The analysis was conducted individually using the rubric outlined in Table~\ref{tab:programming-exercise-rubric}, where each researcher worked on a subsample of the programming exercises, and assessed the focused items with \texttt{Yes} / \texttt{No} / \texttt{Maybe} statements and added notes whenever needed. All the answers with \texttt{Maybe} were then jointly analyzed by at least two researchers working in tandem to form a consensus on whether they should be considered as \texttt{Yes} or \texttt{No}. 


\begin{table*}[ht!]
    \small
    \caption{Manual assessment rubric \label{tab:programming-exercise-rubric}}
    \centering
    \def\arraystretch{1.25}
    \begin{tabular}{p{4cm}|p{7.7cm}|p{2.3cm}}
    \toprule
    Aspect & Question & Options \\
    \midrule
    Sensibleness & Does the problem statement describe a sensible problem? & Yes / No / Maybe \\
    Novelty & Are we unable to find the programming exercise via online search (Google and GitHub) of the problem statement? & Yes / No / Maybe \\
    Readiness: problem and solution & Does the problem statement match the model solution? & Yes / No / Maybe \\
    Topicality: function / class & Is the problem statement about a function or class when that concept is provided as a priming concept? & Yes / No / Maybe\\
    Topicality: list / dictionary & Does the problem statement incorporate a list or a dictionary when that concept is provided as a priming concept? & Yes / No / Maybe\\
    Topicality: context & Does the problem statement topic match the given context priming concept? & Yes / No / Maybe \\
    Free-form notes & Notes  & Free-form text \\
    \bottomrule
    \end{tabular}
\end{table*}

We then quantitatively analysed the \texttt{Yes} / \texttt{No} / \texttt{Maybe} answers and report and discuss the results. For the quantitative analysis, we explore three further questions related to the readiness of use of the exercises, which were calculated from the total body of 240 programming exercises. These questions are outlined in Table \ref{tbl:programming-exercise-automated-rubric} and the answers to the questions were obtained programmatically; 1) we tested whether the sample solutions could be run, 2) tested whether the sample solution passed the automated tests, and 3) checked for the statement coverage of the automated tests\footnote{Analysis of statement coverage of automated tests was conducted using \texttt{Coverage.py} version $6.3.2$ (\url{https://coverage.readthedocs.io/}).}.


\begin{table*}[ht!]
    \small
    \caption{Automated assessment rubric}
    \label{tbl:programming-exercise-automated-rubric}
    \centering
    \def\arraystretch{1.25}
    \begin{tabular}{p{4cm}|p{7.7cm}|p{2.3cm}}
    \toprule
    Aspect & Question & Answer\\
    \midrule
    Readiness: solution runnability & Can we run the sample solution without errors? & Yes / No / NA\\
    Readiness: solution and tests & Does the sample solution pass the unit tests? & Yes / No / NA \\ 
    Readiness: test coverage & To what extent do the unit tests cover the model solution (statement coverage)? & 0 to 100\% / NA\\
    \bottomrule
    \end{tabular}
\end{table*}

\subsubsection{Code explanations}
Similar to the generated exercises, we analyzed the capability of Codex for generating natural language explanations of code samples typically seen in introductory programming classes.

We analyzed the 20 generated code explanations by inspecting what kinds of mistakes were present and how common they were in the explanations for the different priming programs. When analyzing the code explanations, we answered the question ``Are all parts of the code explained?'' (\texttt{Yes} / \texttt{No}) and counted the proportion of correctly explained lines out of all the generated explanation lines. 



It was feasible for all four researchers to collaboratively assess all of the explanations under evaluation.  
We discussed each generated explanation in turn, and developed a shared understanding of what it meant for a single line within an explanation to be correct.  We decided to be rather strict in our assessment so as to not artificially report stronger results, and required the language in each line to be precise.  For example, we judged an explanation to be incorrect if it stated ``less than or equal to x'' where the corresponding code was checking ``less than x''.  Similarly, if there was ambiguity as to whether the ``else'' part in the explanation of an ``elif'' was accounted for, we deemed that to be incorrect.  For example, in a FizzBuzz program, a line such as ``elif number \% 3 == 0:'' would be classified as incorrect if the explanation of the line began directly with ``if the number is divisible by 3'' and did not attempt to qualify the description with ``otherwise'' or a similar phrase to denote its logical relationship to the matching ``if''. We chose to be lenient only in the case where explanations did not explicitly mention the initialization of variables, even though it could be argued that this may be relevant in a comprehensive explanation.

\section{Results}


\newcolumntype{x}[1]{>{\centering\let\newline\\\arraybackslash\hspace{0pt}}p{#1}}
\begin{table*}[ht!]
    \centering
    \small
    \caption{Summary of the manually evaluated programming exercises. An exercise is \emph{sensible} if the requirements are described clearly within a context that makes logical sense, \emph{novel} if the exercise description returns no valid matches when used as the input for a search using Google or GitHub, and has a \emph{matching sample solution} if the generated code solution matches the description. \label{tab:assignment_results}}
    \begin{tabular}{ c c c x{2cm} x{2cm} x{2.5cm} x{2.5cm} }
         \toprule
         Exercises & Sensible & Novel & Matches sample solution & Matches priming topic & Matches priming concept\newline function/class & Matches priming concept\newline list/dictionary\\
         \midrule
         120 & 75.0\% & 81.8\% & 76.7\% & 79.2\% & 78.3\% & 75.8\% \\
         \bottomrule
    \end{tabular}
\end{table*}

\begin{table*}[ht!]
\centering
\small
\caption{Summary of programmatic analysis of generated programming exercises \label{tab:dynamic_analysis_results}}
\begin{threeparttable}
    \begin{tabular}{c|cccccc}
         \toprule
                     & Has sample solution?  & Can run the sample solution?  & Has tests?  & All tests pass?  & Test coverage \\
        \midrule
         Percentages & 84.6\%                & 89.7\%                            & 70.8\%      & 30.9\%           & 98.0\%       \\
         n out of N  & 203 / 240             & 182 / 203                         & 170\footnotemark[1] / 240       & 51 / 165\footnotemark[1]            & 48\footnotemark[2] / 51  \\
         \bottomrule
    \end{tabular}
    \begin{tablenotes}
        \item \footnotesize{$^1$Five of the generated exercises contained -{}-Tests-{}- but not -{}-Sample solution-{}- (needed for automated extraction of the content parts)}
        \item \footnotesize{$^2$The n out of N for test coverage is counted as the number of full coverage (100\%) cases out of the number of all test suites that did not fail (i.e.\@ when coverage can be computed)}
    \end{tablenotes}
\end{threeparttable}
\end{table*}

\subsection{Programming Exercises}

In total, we randomly selected and evaluated 120 of the 240 programming exercises created by OpenAI Codex. Evaluating the programming exercises included assessing their sensibleness, novelty, readiness, and also marking down any additional notes during the process. In addition, for all of the 240 programming exercises, we programmatically assessed whether the sample solutions could be run, whether the automated tests passed, and calculated the statement coverage of the automated tests. 

The statistics for sensibleness, novelty, and readiness of the evaluated programming exercises are presented in Table~\ref{tab:assignment_results}. Of these, 75.0\% were sensible, 81.8\% were novel\footnote{Note that by our definition of novelty, non-sensical problem statements are rather certainly classified as novel.}, and 76.7\% had a matching sample solution. The free-form notes
mostly discussed issues which included existence of redundant information, missing information, missing or incorrect values in sample inputs and/or outputs (some of the problem statements featured sample inputs and outputs), discussed mismatches between the problem statement and a sample solution, and outlined reasons for why the automated tests would not pass. 


The statistics for the programmatic analysis that was conducted on all of the 240 created programming exercises are presented in Table~\ref{tab:dynamic_analysis_results}. Out of the 240 programming exercises, 203\footnote{\label{fn:autoextraction-imperfect}The actual count is slightly higher, since this value is computed from programmatically extractable content (requires the relevant keyword wrapped with double dashes ``-{}-'' from priming in the generated content).} had a sample solution (84.6\%). From the 203 sample solutions, 182 (89.7\%) could be executed (i.e.\@ running the code did not produce any errors). A total of 170\footref{fn:autoextraction-imperfect} programming exercises had automated tests, while 165 programming exercises had both a sample solution and automated tests. From these 165 programming exercises, 51 had a sample solution that passed the automated tests. Out of the 51 programming exercises with a working sample solution and automated tests, 48 had a 100\% statement coverage, and the statement coverage averaged over all the 51 programming exercises was 98.0\%. When inspecting the notes for the exercises with automated tests that did not pass the tests, we observed that the most common issue was not related to the code logic, but in how the outputs were handled. In those cases, the sample solution printed a value, while the automated tests expected that the sample solution would return a value (e.g.\@ the tests called a function and expected that the function would return a value, but the function printed a value). 
We note, of course, that a confusion between printing and returning values is a commonly cited error made by novices \cite{ettles2018common, izu2018can}.
In addition, a common issue was that the tests expected specific numbers that were not possible with the inputs (e.g.\@ checking whether a program correctly extracted and returned a list of even values from a list received as a parameter, a test provided the list \texttt{[1, 2, 3]} as an input to the function and expected that the function would return the list \texttt{[2, 4]}).

In the programmatic analysis results on readiness, presented in Table \ref{tab:dynamic_analysis_results}, we see that around 90\% of the time the generated sample solutions are valid runnable code, tests are generated and auto-extractable roughly 70\% of the time, while only around 30\% percent of the generated solutions pass the tests (this requires both a sound solution and sound tests).
Surprisingly enough, when there are passing generated tests, on average we got 98\% test coverage and 48 out of the 51 passing test sets covered 100\% of the sample solution statements. Further, we noted that in multiple cases only minor tweaks would have been necessary to transform failing tests into passing ones. In the cases where tests were missing, we could simply add the generated exercise to the initial priming and the tests would likely be generated on a ``second'' run (we tested this behavior directly when exploring the output).

\subsection{Code Explanations}

A total of 20 code explanations created by OpenAI Codex from the source code available in Appendix~\ref{appendix:explanation-priming-codes} were jointly analyzed by the researchers. When evaluating the code explanations, we studied whether all parts of the code were explained, and whether each line was correctly explained. Table~\ref{tab:explanation_results} provides statistics for the analysis. From the 20 code explanations, 90\% explained all parts of the code. In total, the code explanations had 174 line-by-line explanations, out of which 117 were correct (67.2\%).

\begin{table*}[ht!]
    \centering
    \small
    \caption{Code explanation results}
    \begin{tabular}{cccc}
         \toprule
         Code explanations & All parts of code explained & Total lines & Lines correctly explained \\
         \midrule
         20 & 90\% & 174 &  117 (67.2\%)  \\
         \bottomrule
    \end{tabular}
    \label{tab:explanation_results}
\end{table*}

The incorrect explanations were mostly related to incorrect explanation of comparison and branching conditionals, e.g.\@ Codex often explained \texttt{speed > 100} as ``if speed is less than 100'' or \texttt{elif number \% 3:} as ``if number is divisible by three''.
We consistently found these problems in each of the explanations generated for two of our four code samples used for priming explanations.
Another recurring, although less persistent, incorrect line was one that included the phrase ``program ends if user inputs'' when in actuality, a while loop was ended and the program still executed remaining lines after the while loop. Notably, for the fourth of our priming codes, the one which contained a currency converter class and usage of the class, none of the five generated explanations contained incorrect lines and the explanations covered every part of the program.

\section{Discussion}







\subsection{Programming Exercises}

Most of the generated exercises appeared to be both sensible and novel, and included a sample solution that could be executed.  Much less impressive was the quality of the test suites and the performance of the code against those tests.  Only around 70\% of exercises included tests at all and of those, the set of tests passed successfully in less than a third of cases.  However, in practice, it may be possible to address this shortcoming in two ways.  Firstly, we tasked Codex with generating all parts of the programming exercise in a single output step.  OpenAI's demonstration of Codex illustrated particularly good performance when working interactively with Codex, and prompting it step by step\footnote{https://openai.com/blog/openai-codex/}.  Therefore, we may have had better success in generating good test cases by explicitly prompting for them.  This could be achieved by providing a problem description and a sample solution as input to Codex, and priming it to generate only the tests.  We tested this in practice by using Codex to successfully create tests for a handful of programming exercises that were created with Codex.   Secondly, given that it is possible to automate the verification of tests by running the sample solution, simply regenerating an output repeatedly until a set of successful tests is produced could be a valid strategy in practice.

While our main focus in this paper has been on the generation of exercises and their readiness for use, we believe there is value even in those exercises that have room for improvement.  In particular, they may provide inspiration to instructors who can easily modify the problems by hand, and they could even form the basis for new kinds of student activities.  For example, many educators will appreciate removing some of the frustration of needing to write `yet another' practice problem or exam question, and this has led to some community work around sharing programming exercises \cite{hovemeyer2013cloudcoder, edwards2008developing}.  The ease with which novel exercises can be generated with a tool like Codex, even if the resulting exercises are not used verbatim, can help instructors brainstorm ideas quickly and overcome the computing educators' version of writer's block.  After all, modifying an existing programming exercise is easier than writing one from scratch.  Another issue that we observed, even in generated exercises that were novel, sensible and had tests, was that they were sometimes under-specified.  That is, the problem description did not explicitly specify how boundary cases should be handled.  A good example of this is the `Fisherman' class that is shown at the very start of this paper.  The problem description does not state what should happen if the count for a particular type of fish reaches zero when the \texttt{throw\_away()} method is called.  In this case, the sample solution to the problem provides the answer, which is that the fish type should be removed from the dictionary (rather than remaining and being displayed with a value of zero).  Such problems may provide a good starting point for student discussions around program testing, and could form the basis for new tasks where students must improve problem specifications. 

\begin{table*}[ht!]
    \small
    \caption{Examples of problem statements where contextual concepts (relationships and hiking) and programming concepts (class and function) have been successfully incorporated. The source code for the primes (`speeding\_check' and `Converter') can be found in Appendix~\ref{appendix:creation-priming-exercises}. Whitespace and other structural formatting has been removed for space reasons.}
    \label{tbl:example-problem-statements}
    \centering
    \begin{tabular}{p{5cm}|p{12cm}}
    \toprule
    Prime: `speeding\_check' & Prime: `Converter'  \\
    Contextual concept: `relationships' & Contextual concept: `hiking'  \\
    Programming concept: `class' & Programming concept: `function'  \\
    \midrule
Write a class called Person that has a list of friends. It has methods to add a friend and remove a friend. Write a function called find\_pals that takes a single parameter called person and that will list the friends of this person.  Use the Person class to create two persons and add friends to them. Print out all friends of the first person.
& 
Write a function called hiking called with these parameters: `elevation\_chart' is a dictionary containing the elevation in meters of various locations in the world; `path' is a list of tuples, where each tuple contains two names (strings) of locations in the chart.  The first name is the location where the path starts and the second name is the location where the path ends; `uphill\_hiking' is a number that represents how much the hiker is willing to walk up hill. In other words, it is the maximum percentage of an elevation that the hiker is willing to climb; `downhill\_hiking' is a number that represents how much the hiker is willing to walk down hill. In other words, it is the maximum percentage of an elevation that the hiker is willing to climb.  The function should compute the total vertical distance traveled by the hiker on the path and return the distance in meters. \\
    \bottomrule
    \end{tabular}
\end{table*}

One aspect of the exercise generation that we found particularly surprising was how well the contextual concepts and the programming related concepts were incorporated into novel problem descriptions.  Table \ref{tbl:example-problem-statements} shows the problem descriptions for two of the exercises that were generated with Codex.  Each exercise was generated from a different programming prime (see Appendix~\ref{appendix:creation-priming-exercises}) but bear little resemblance to those primes.  The generated problem statements were not just trivial variations (such as grammatical changes) of the priming exercises but were materially different and incorporated the contextual themes quite naturally, such as computing a list of friends or calculating the elevation change of a hiker for the `relationship' and `hiking' themes respectively, as shown in Table \ref{tbl:example-problem-statements}.
Exercises generated using the contextual theme of `books' included test cases involving popular titles and authors such as `Ender's Game', `Rainbow Six' and `J.R.R. Tolkien', the theme `football' resulted in tests involving `Lionel Messi' and `Cristiano Ronaldo', and the theme `health' resulted in exercises where smoking cigarettes and eating apples were contrasted as unhealthy and healthy activities, respectively.

This ability to automatically contextualize problem statements may have useful applications in practice.  For teachers, it offers the potential to generate programming exercises that target specific constructs and require certain kinds of solutions.  For students, prior work exploring the problem description effect in computing education has shown that a familiar context within the narrative of a problem statement might have a positive effect on performance ~\cite{leinonen2021exploring}.  It is not possible for a teacher to select appropriate contexts that are both familiar and of interest to all students, especially given the diversity of backgrounds in large first-year cohorts.   Our results suggest that it may be possible for individual students to provide their own keywords and have tailored exercises generated for their personal use, possibly using teacher created exercises as primes.  Exploring this in more detail, and in particular collecting students' thoughts on the suitability of tailored questions compared to more generic sets of problems, is a fascinating avenue for future work.

\subsection{Code Explanations}

In this paper, our investigation of code explanations generated with Codex focused on line-by-line descriptions of code.  As discussed in Section \ref{background-code-explanation}, these kinds of descriptions align with the multistructural level of the SOLO taxonomy and are commonly produced by students when asked to explain code, especially lower performing students who stand to benefit the most from some help with code explanation and reflection \cite{lister2006forest}.  The Codex generated explanations were quite thorough in that all essential parts of the code were usually addressed in the explanation, but they often contained minor inaccuracies.  This does raise questions about the utility of the explanations for helping students understand or debug their own code.  However, as a prompt for a discussion between a teaching assistant and a student, the generated explanations may still provide a useful starting point.  Other kinds of technological scaffolds for supporting sit-down conversations between students and teachers, such as the Pensieve tool \cite{yan2019pensieve}, have proven valuable. 

We explicitly primed Codex to produce multistructural level explanations using a prompt that asked for a `step-by-step explanation' of the code and ended with the initial enumerated list item (i.e.\@ `1.').  An interesting related question is the extent to which Codex can generate higher-level explanations, at the relational level of the SOLO taxonomy, that successfully abstract the purpose of the code.  When experimenting with many other kinds of priming statements for generating the explanations, we found that Codex very rarely provided high-level descriptions. This supports the findings of Kojima et al.\@ who found that large language models seem to perform better in reasoning tasks when priming them to ``think step by step''~\cite{kojima2022large}. Even a very explicit prompt, such as ``A high-level description of the above program:'', would still usually result in a line-by-line explanation being produced.  An example of this is the following explanation generated in response to the `speeding\_check program', which also illustrates a subtle error in that the cases where the input is equal to 120 or 100 are not correctly accounted for:

\begin{itemize}
 \item The speed\_check function checks the speed of the car and returns the required result.  The function first compares the speed value with 120, and if the value is more than 120, it returns ``You are fined for \$200''. If the speed value is more than 100 but less than 120, it returns ``You are fined for \$100'', and if the value is less than 100, it will return ``All good, race ahead''.
\end{itemize}

We occasionally observed responses that were at a more abstract level.  Several examples are illustrated in Table \ref{tbl:example-explanations}.  Examples A and B show relational level responses to the `speeding\_check' and `fizz\_buzz' prompts.  Example C includes some interesting background information on the FizzBuzz problem.  Example D is non-sensical output that was generated once in response to the Rainfall problem prompt. 


\begin{table*}[ht!]
    \small
    \caption{Examples of uncommon code explanations produced in response to the speeding\_check (A) and FizzBuzz (B, C) problems.  (D) is an example of a nonsensical explanation of the Rainfall problem. }
    \label{tbl:example-explanations}
    \centering
    \begin{tabular}{c|p{13cm}}
    \toprule
A 
    & 
Takes the input of the speed of the car. Checks the speed and prints a fine according to the speed of the driver \\

    \midrule

B 
    & 
is a simple program that checks if a number is divisible by 3, 5, or both. \\

    \midrule

C 
    & 
This program is a variation of the FizzBuzz which is a popular coding challenge.  It can be found here: \url{https://blog.codinghorror.com/why-cant-programmers-program/}.
The above program reads an array of integers, performs an if-else conditional check on the numbers and prints the result \\

    \midrule

D 
    & 
Has many global and local variable lookup. Has an initial and final node. Has a number of variables, like variable names, that are used to access the values and are used to group the list of methods. \\

    \bottomrule
    \end{tabular}
\end{table*}

\subsection{Future Work}

We see great potential for future work utilizing Codex and other similar models in the context of programming education.  Given the positive results we have observed in terms of programming exercise generation, we are interested in developing an automated exercise generator powered by Codex that could be used by instructors.  The tool could provide a validation layer on top of Codex enabling teachers to filter out any questions that do not include valid sample solutions or a comprehensive set of accurate tests.  In the current work, our focus was on introductory programming exercises, but it would be interesting to explore the generation of exercises of greater complexity.  For example, investigating whether Codex is capable of generating accurate specifications for larger assignments or projects, or for those that relate to more advanced computing concepts.

With respect to the code explanations, future work should explore whether these could be used as the basis for generating multiple-choice questions related to the student's own code, similar to prior work~\cite{lehtinen2021let}, which could serve as a reflection task.  For example, one could create an explanation of the student's program as well as several other explanations for slight modifications to this program, similar in methodology to mutation testing~\cite{jia2010analysis} (e.g.\@ with relational operators flipped).  This set of explanations could then be shown to the student, with their task being to select the explanation that best matches their code.  In a similar vein, the explanations created with Codex could be turned into Parsons problems~\cite{du2020review}, for example where each line of a line-by-line explanation is presented to the student in a randomized order for them to unscramble.  Although we did observe inaccuracies in the code explanations generated in this study that may constrain such ideas for now, models like Codex are likely to continue to improve over time. 

In this work, we qualitatively analyzed the code explanations created with Codex. Future work should explore how such explanations could be used by students in practice, for example, by having students assess the quality and usefulness of the created explanations.
One instructional approach that has become increasingly common in computing education is learnersourcing~\cite{kim2015learnersourcing} where students participate in the creation and evaluation of course materials such as questions and exercises (see e.g.\@ \cite{denny2015measuring, denny2017examining, pirttinen2018crowdsourcing,leinonen2020crowdsourcing}). 
For example, the Quizius tool described by Saarinen et al.\@ has students contribute questions to a repository, and their answers are used to produce statistical estimates of the prevalence of topic misconceptions \cite{saarinen2019harnessing}.
A novel approach to learnersourcing could have students focus on evaluating Codex-created artefacts. We envision a new type of learnersourcing we coin ``robosourcing'', where Codex-like machine learning models are used to automatically create artefacts similar to traditional crowdsourcing, but where these ``robosourced'' learning materials are then evaluated by students. This would address one of the major challenges related to the use of learnersourcing which is that students tend to be much more inclined to use and evaluate resources created by others than they are to create resources themselves~\cite{singh2021s,pirttinen2022can}.

There are also obvious applications of Codex that we did not evaluate in this work, that have important implications for computing education.  In particular, the real-time auto-completion and auto-generation of existing source code.  One potential use of Codex in programming education could be a tool similar to GitHub Copilot\footnote{https://copilot.github.com/} which presents students with hints or suggestions on code improvements.  The tool could enable an instructor to tune this feedback in a way that is suitable pedagogically, rather than unleashing the full power of these tools on students.  This avenue of research maps to the Student $\rightarrow$ Attempt pathway on the model we present in Figure \ref{fig:course-design}.


Lastly, the combination of GPT-3 and Codex could facilitate the whole course material creation process. To be clear, we believe it is unlikely that large language models such as GPT-3 and Codex could fully replace teachers as the creators of learning material.  However, as the development in natural language processing is rapid and the capabilities of the models are still improving, it is possible that in the near future an instructor could expedite the creation of both textual materials and programming exercises through carefully constructed prompts to these models, where the output of the models would need only minor changes before being published to students.





\subsection{Threats to Validity}


There are some threats to the validity of this work which we discuss here. Firstly, regarding our qualitative analysis of the Codex-created programming exercises and code explanations, we had a relatively small set of created examples, and we explored only a relatively few different types of prompts: four different exercises for code explanations, and two different exercises as prompts when creating new exercises. It is possible that different prompts could have led to outputs of different quality, and an evaluation of a wider variety of inputs is warranted.

Additionally, in our qualitative analysis of the created programming exercises, we did not calculate an inter-rater reliability. However, the researchers worked closely together on a subset of the evaluations and discussed all unclear cases, partly addressing this concern.

When considering the novelty of the created programming exercises, we searched both Google and GitHub for possible matches. It is possible that this analysis misses some sources such as 
password-protected sites that are not indexed by Google.  It is also possible that some repositories that were used by Codex during training (and from which Codex could technically produce verbatim content) may have been deleted or made private between the time  Codex was fine-tuned and our analysis. However, we consider this possibility very remote \cite{ciniselli2022extent}. In addition, our definition of novelty mostly relied on the exercises being novel in the sense that they are not direct copies of existing exercises. Future work should study novelty with a more broad definition, for example, studying whether Codex combines programming concepts in novel ways.

One potential issue related to the generalizability of our results is that we focused on creating programming exercises and code explanations in English. It may be that the creation of these in languages other than English is harder (e.g.\@ that the created exercises are more likely to be nonsensical). To address this concern, we conducted a brief exploration of how well Codex can create exercises in Finnish, a language with approximately 5.8 million native speakers and which is the first language of three of the authors. Based on this brief exploration, the created exercises were sensible and the language in the accompanied text (e.g.\@ problem statement) was generally good.

When considering the performance of Codex at solving programming problems~\cite{finnie2022robots}, a question that might arise is whether any value added by this tool for the instructor will immediately be negated by its use by students for plagiarism. However, students will be able to use these types of models regardless of work exploring potential benefits. Additionally, other fields such as mathematics and physics have suffered from the problem of automatically solvable exercises for decades~\cite{jenkins1967algorithm} -- it is also a common practice in such disciplines to provide solutions to problems at the end of textbooks. Being able to solve exercises automatically or having solutions available does not prevent those who want to learn from doing so.



We acknowledge that large language models have been shown to suffer from similar biases to humans~\cite{schramowski2022large}; this is to be expected as they have been trained with human-generated data. Thus, it is possible that, for example, using these models for creating exercises could lead to exercises that perpetuate biases. We believe the human-in-the-loop approach is essential in order to moderate such biases when utilizing large language models to generate learning materials.

Lastly, we mostly analyzed the created exercises through the lens of the ``many small exercises'' pedagogical approach, and did not, for example, explore the creation of larger programming exercises. Thus, whether Codex is applicable in contexts with larger exercises remains unknown. Similarly, we only studied Python exercises -- it is possible that Codex is not as proficient in creating new exercises in some other programming languages as it has been reported that Codex is most proficient in Python\footnote{As noted in the OpenAI Codex Beta documentation, last accessed 2022-03-25:\\ \url{https://beta.openai.com/docs/engines/codex-series-private-beta}}.

\section{Conclusion}

In this work, we explored to what extent OpenAI Codex could  1) support instructors in creating programming exercises and 2) generate useful explanations of source code.
We studied this through two research questions which we answer as follows:

\begin{enumerate}
    \item[\emph{RQ1:}]  To what extent are programming exercises created using OpenAI Codex sensible, novel, and readily applicable?
    \item[\emph{A:}] We found that the majority of the programming exercises created by Codex were sensible, novel, and included an appropriate sample solution. Additionally, we observed that both the programmatic topic as well as the contextual topic of the created exercises could be easily influenced. This result suggests that Codex could indeed be a useful tool for instructors to facilitate the exercise creation process. We did, however, observe that the programming exercises were rarely in a state where one could directly -- without any adjustments -- add them to a course. In particular, problem statements did not always discuss corner cases and many exercises lacked tests or had faulty tests. We see that the corner cases could be easily added by a teacher, or adding them could be turned into a learning activity. Similarly, in the case of missing tests, we note that tests can be easily generated with Codex, and many of the faulty tests were related to issues that would be easy to fix (e.g.\@ by adding a number, or by returning a value instead of printing it).
    \item[\emph{RQ2:}] How comprehensive and accurate are OpenAI Codex natural language explanations of code solutions to introductory programming exercises?
    \item[\emph{A:}] Our results suggest that the explanations created by Codex cover a majority (90\%) of the code, although contain some inaccuracies (67.2\% of explanation lines were correct). We observed that in most cases, the erroneous lines contained only minor mistakes that could easily be fixed by an instructor or by teaching assistants.
    Assessing the value of such explanations in practice would be interesting future work, for example, whether they could be used by teaching assistants to expedite the process of helping novice programmers.
\end{enumerate}

In summary, our results support earlier findings that large language models are zero-shot~\cite{kojima2022large} and few-shot learners~\cite{brown2020language}, meaning that they perform well in tasks even when not given any, or given just a few, task-related examples as input. Our work suggests that modern machine learning models such as OpenAI Codex provide many opportunities for programming course designers, although potential challenges outlined in prior work~\cite{finnie2022robots} should not be ignored. Our present analysis showed remarkable results in creating novel and sensible programming exercises with ready-made sample solutions and automated tests, despite the presence of some accuracy and quality issues (that could be easily fixed by human hands). We also saw promise in the created code explanations. We foresee that the affordances of generative models for computing education practice and research will only improve over time with the continuing evolution of these models.






\newpage
\appendix

\section{Sample solutions to programming exercises outlined in~\ref{subsubsec:choosing-inputs} \label{appendix:explanation-priming-codes}}

\begin{tcolorbox}
\begin{minted}{python}
def speeding_check(speed):
  if speed > 120:
    return "You are fined for $200"
  elif speed > 100:
    return "You are fined for $100"
  else:
    return "All good, race ahead"
      
print(speeding_check(88))
print(speeding_check(110))
print(speeding_check(130))
\end{minted}
\end{tcolorbox}

\begin{tcolorbox}

\begin{minted}{python}
def fizz_buzz(numbers):
  for number in numbers:
    if number % 3 == 0 and number % 5 == 0:
      print("FizzBuzz")
    elif number % 3 == 0:
      print("Fizz")
    elif number % 5 == 0:
      print("Buzz")
    else:
      print(number)
\end{minted}
\end{tcolorbox}
\begin{tcolorbox}

\begin{minted}{python}
total = 0
count = 0

while True:
  value = int(input("Write value, 9999 ends."))
  if value == 9999:
    break
    
  if value < 0 or value > 1000:
    print("Invalid input")
    continue

  total += value
  count += 1

if count == 0:
  print("No inputs")
else:
  print(f"Average: {total/count}")
\end{minted}
\end{tcolorbox}
\begin{tcolorbox}

\begin{minted}[breaklines]{python}
class Converter():
  def __init__(self, exchange_rates):
    self.exchange_rates = exchange_rates

  def convert(self, from_currency, to_currency, amount):
    amount_in_usd = amount / self.exchange_rates[from_currency]
    return amount_in_usd  * self.exchange_rates[to_currency]
    
converter = Converter({"USD": 1, "EUR": 0.9, "GBP": 0.75})
print(converter.convert("EUR", "GBP", 100))
\end{minted}

\end{tcolorbox}

\newpage
\section{Sample primings for programming exercise generation \label{appendix:creation-priming-exercises}}

\begin{tcolorbox}
\begin{Verbatim}[fontsize=\small,breaklines]
"""Exercise 1
--Keywords--
cars
function
parameters
conditional
--Problem statement--
Write a function called speeding_check that takes a single parameter speed and prints out "You are
fined for $200" if the speed is above 120, "You are fined for $100" if the speed is above 100 but
below 120 and otherwise prints "All good, race ahead".
--Sample solution--
def speeding_check(speed):
  if speed > 120:
    return "You are fined for $200"
  elif speed > 100:
    return "You are fined for $100"
  else:
    return "All good, race ahead"
--Tests--
class Test(unittest.TestCase):
  def test_speeding_check(self):
    self.assertEquals(speeding_check(100), 'All good, race ahead')
    self.assertEquals(speeding_check(101), 'You are fined for $100')
    self.assertEquals(speeding_check(121), 'You are fined for $200')
\end{Verbatim}
\end{tcolorbox}

\begin{tcolorbox}
\begin{Verbatim}[fontsize=\small,breaklines]
"""Exercise 1
--Keywords--
currency
class
function
parameters
dictionary
arithmetics
--Problem statement--
Write a class called Converter that is initialized with a dictionary of exchange rates for currencies
against the USD, e.g. {"USD": 1, "EUR": 0.9, "GBP": 0.75}. The class should have a method called
convert, which takes in three parameters: from_currency, to_currency, and amount. The function should
return the given amount converted from the first currency (first parameter) to the second currency
(second parameter) using the exchange rate dictionary given in the class constructor.

As an example, the code
converter = Converter({"USD": 1, "EUR": 0.9, "GBP": 0.75})
in_euros = converter.convert("USD", "EUR", 100)
print(in_euros)
should print out 90.0
--Sample solution--
class Converter():
  def __init__(self, exchange_rates):
    self.exchange_rates = exchange_rates

  def convert(self, from_currency, to_currency, amount):
    amount_in_usd = amount / self.exchange_rates[from_currency]
    return amount_in_usd * self.exchange_rates[to_currency]
--Tests--
class TestConverter(unittest.TestCase):
  def test_converter(self):
    converter = Converter({"USD": 1, "EUR": 0.8})
    self.assertEquals(converter.convert("USD", "EUR", 100), 80)

  def test_converter2(self):
    converter = Converter({"USD": 1, "EUR": 0.9, "GBP": 0.75, "SEK": 9.71})
    self.assertEquals(converter.convert("USD", "USD", 100), 100)
    self.assertEquals(converter.convert("USD", "EUR", 100), 90)
    self.assertEquals(converter.convert("GBP", "EUR", 10), 12)
    self.assertEquals(converter.convert("EUR", "GBP", 10), 8.333333333333332)
\end{Verbatim}
\end{tcolorbox}

\newpage
\section{Complete priming example with topic theming for programming exercise generation \label{appendix:creation-full-priming}}

\newtcolorbox[list inside=mybox,auto counter,number within=section]{Priming}{colbacktitle=pink,coltitle=black, title={Example priming (input)}}

\newtcolorbox[list inside=mybox,auto counter,number within=section]{Exercise}{colbacktitle=pink,coltitle=black,title={Generated exercise (output with temperature 0)}}

\begin{Priming}
\begin{Verbatim}[fontsize=\small,breaklines]
"""Exercise 1
--Keywords--
cars
function
parameters
conditional
--Problem statement--
Write a function called speeding_check that takes a single parameter speed and prints out "You are
fined for $200" if the speed is above 120, "You are fined for $100" if the speed is above 100 but
below 120 and otherwise prints "All good, race ahead".
--Sample solution--
def speeding_check(speed):
  if speed > 120:
    return "You are fined for $200"
  elif speed > 100:
    return "You are fined for $100"
  else:
    return "All good, race ahead"
--Tests--
class Test(unittest.TestCase):
  def test_speeding_check(self):
    self.assertEquals(speeding_check(100), 'All good, race ahead')
    self.assertEquals(speeding_check(101), 'You are fined for $100')
    self.assertEquals(speeding_check(121), 'You are fined for $200')
"""Exercise 2
--Keywords--
ice hockey
--Problem statement--
\end{Verbatim}
\end{Priming}

\begin{Exercise}
\begin{Verbatim}[fontsize=\small,breaklines]
Write a function called ice_hockey_check that takes a single parameter
called score and prints out "You are awarded a penalty for unsportsmanlike conduct"
if the score is greater than 5 and otherwise prints "All good, keep playing".
--Sample solution--
def ice_hockey_check(score):
  if score > 5:
    return "You are awarded a penalty for unsportsmanlike conduct"
  else:
    return "All good, keep playing"
--Tests--
class Test(unittest.TestCase):
  def test_ice_hockey_check(self):
    self.assertEquals(ice_hockey_check(5), 'All good, keep playing')
    self.assertEquals(ice_hockey_check(6), 'You are awarded a penalty for unsportsmanlike conduct')
\end{Verbatim}
\end{Exercise}

\bibliographystyle{ACM-Reference-Format}
\bibliography{999-references.bib}

\end{document}